\documentclass[final,5p,twocolumn,12pt]{elsarticle}

\usepackage{graphicx}
\graphicspath{{fig/}}
\usepackage{epsfig}
\usepackage{tabularx}

\usepackage{units}
\usepackage{amssymb}
\usepackage{color}
\usepackage{lineno}
\usepackage[highlight]{neel}

\usepackage{hyperref}
\definecolor{link_green}{rgb}{0.0,0.7,0.0}
\hypersetup{pagebackref,breaklinks,citecolor=link_green,colorlinks,bookmarks,bookmarksopen,
  pdfauthor="Olivier Fruchart",pdfstartview=FitH}
\usepackage[all]{hypcap}

\journal{Journal of Magnetism and Magnetic Materials}

\begin{document}

\begin{frontmatter}



\title{Tunable magnetic properties of arrays of Fe(110) nanowires grown on  kinetically-grooved W(110) self-organized templates}


\author[neel]{B. Borca\fnref{addrBogdana}\corref{corrBogdana}}
\ead{bogdana.borca@uam.es}
\tnotetext[addrBogdana]{Present address: Departamento de F\'isica de la Materia Condensada, Universidad Aut\'onoma de
Madrid, Cantoblanco 28049, Madrid, Spain}
\cortext[corrBogdana]{Principal corresponding author}

\author[neel]{O. Fruchart\corref{corrFruche}}
\ead{Olivier.Fruchart@grenoble.cnrs.fr}
\cortext[corrFruche]{Permanent corresponding author}

\author[neel,spintec]{E. Kritsikis}
\author[neel]{F. Cheynis\fnref{addrFabien}}
\tnotetext[addrFabien]{Present address: Centre Interdisciplinaire de Nanoscience de Marseille (CINaM), CNRS - UPR 3118 Campus de Luminy Case 913, 13288 Marseille Cedex 9, France}

\author[neel]{A. Rousseau\fnref{addrAnthony}}
\tnotetext[addrAnthony]{Present address: Institut de Chimie de la Mati\`{e}re Condens\'ee de Bordeaux (ICMCB), CNRS - UPR 9048, Avenue du Docteur Schweitzer 87, 33608 PESSAC Cedex, France}

\author[neel]{Ph. David}
\author[neel]{C. Meyer}
\author[neel]{J. C. Toussaint}

\address[neel]{Institut N\'eel (CNRS and UJF), 25 rue des Martyrs, BP 166, F-38042 Grenoble, France}
\address[spintec]{Spintec (CEA, CNRS, INPG, UJF, 17 rue des Martyrs, F-38054 Grenoble, France}

\begin{abstract}
We report a detailed magnetic study of a new type of self-organized nanowires disclosed briefly previously~[B. Borca \etal, Appl. Phys. Lett. \textbf{90}, 142507 (2007)]. The templates, prepared on sapphire wafers in a kinetically-limited
regime, consist of uniaxially-grooved W(110) surfaces, with a lateral period here tuned to \thicknm{15}. Fe deposition leads to the formation of (110) \thicknm{7}-wide wires located at the bottom of the
grooves. The effect of capping layers (Mo, Pd, Au, Al) and underlayers (Mo, W) on
the magnetic anisotropy of the wires was studied. Significant discrepancies with figures known for thin
flat films are evidenced and discussed in terms of step anisotropy and strain-dependent surface anisotropy. Demagnetizing coefficients of cylinders with a triangular isosceles cross-section have also been calculated, to estimate the contribution of dipolar anisotropy. Finally, the dependence of magnetic anisotropy with the interface element was used to tune the blocking temperature of the wires, here from $\unit[50]{K}$ to $\unit[200]{K}$.
\end{abstract}


\end{frontmatter}

\def\Kin{K_\mathrm{in}}%Constante d'anisotropie dans le plan
\def\Kout{K_\mathrm{out}}%Constante d'anisotropie hors du plan

\section{Introduction}
\label{intro}

Bottom-up processes are subject to an increasing interest for the fabrication of nanostructures at surfaces.
These processes are inexpensive alternatives to the ever-rising cost of lithography with the decreasing pitch.
Self-organized dots or wires \ie arranged as a regular array on surfaces, are of particular interest because a
narrow size distribution may arise as a consequence of the order\cite{Weiss}. Such studies are motivated by the
demand for an ever higher density of hard-disk drives, for which goal reducing the dispersion of grain size and
magnetic properties of the media is a requirement. Beyond this technological motivation self-organized systems
contribute to the fundamental understanding of magnetism in low dimensions, based on chemical\cite{Fer99}
or epitaxial routes\cite{Fru05}. Nanostructures from the micron size\cite{Jubert, Bode84, Wachowiak02} down to
the atomic size \cite{Weiss, Gambardella, Gam03, Repp} have been studied, for studying in low-dimensions \eg
the micromagnetics of magnetic domains and domain walls, and magnetic moments and anisotropy, respectively.

We focus here on the fabrication and magnetic properties of self-organized arrays of epitaxial nanowires.
This system addresses the tuneability in terms of period and magnetic properties, which are one of the prerequisites for such systems to be relevant for applications.
Magnetic wires at surfaces are often achieved by step-decoration of
vicinal surfaces\cite{Hauschild1998,Dallmeyer,Fruchart04}. In this approach a crystal has to be polished with a
specific miscut to achieve the desired period. Templates resulting from kinetic effects are potentially more
versatile as the period can be changed with parameters like temperature. Ion etching under grazing incidence may
be used to create ripples and wires on surfaces\cite{Val02}. However a significant control of the period has not
been demonstrated for magnetic materials so far. Here we report on the advancement of a recent alternative
approach \cite{borca1,Yu}, based on the \textsl{growth} of a body-centered-cubic $\mathrm{W}(110)$ template on
nominally-flat sapphire, instead of etching. Fe wires are then grown at the bottom of the trenches in a
layer-by-layer fashion. Using the dependance of surface magnetic anisotropy with the interfacial element, we
could tune the magnitude of the magnetic anisotropy of the wires grown on such templates using suitably-chosen
capping layers and underlayers. As a consequence this gives us a mean to tune the mean blocking temperature of
the wires, here from $\tempK{50}$ to $\tempK{200}$.

% \appendix

\section{Experimental procedures}
% \label{}
The samples were grown in a set of ultra-high vacuum chambers, using pulsed-laser deposition with a Nd-YAG laser
operated at $\unit[532]{nm}$. The
metallic films and nanostructures are grown starting from commercial sapphire single crystals. The typical fluence is $\unit[1-3]{J.cm^{-2}.pulse^{-1}}$ depending on the element, yielding a
rate of deposition around $\unit[1]{\AA/min}$. The deposition chamber is equipped with a quartz microbalance, sample heating and a
translating mask for the fabrication of wedge-shaped samples. We use a Riber $\unit[10]{keV}$ Reflection High
Energy Electron Diffraction (RHEED) setup with a CCD camera synchronized with laser shots, which thus permits
operation during deposition. An Omicron room-temperature Scanning Tunneling Microscope (STM), used in the
constant current mode, and  an Auger Electron Spectrometer (AES) with a MAC-II analyzer are located in connected chambers. A quantitative analysis was performed based on the energy range $\unit[100-800]{eV}$ and an incident beam energy of $\unit[5]{keV}$, and focusing on the $NOO$ and $LMM$ transitions of W and Fe, respectively centered around $\unit[180]{eV}$ and $\unit[650]{eV}$. Each measured spectrum was first derived, then smoothed and normalized by the background, representing the signal of secondary electrons detected in the energy range where no characteristic peaks were expected. %The position of energy peaks of both reference spectra ( of W and Fe respectively) could be shifted manually to account for AES analyzer energy drift before the fitting procedure. Typical energy shifts of \unit[???]{eV} were used.\comment{BB:look for the value}.
A detailed description of the chambers and
growth procedures can be found in \cite{Fru}. The magnetic measurements were performed \exsitu
with a Quantum Design MPMS Superconducting QUantum Interference Device (SQUID) magnetometer.

\section{Nanowires preparation}

\begin{figure}
    \center
    \includegraphics[width=77mm]{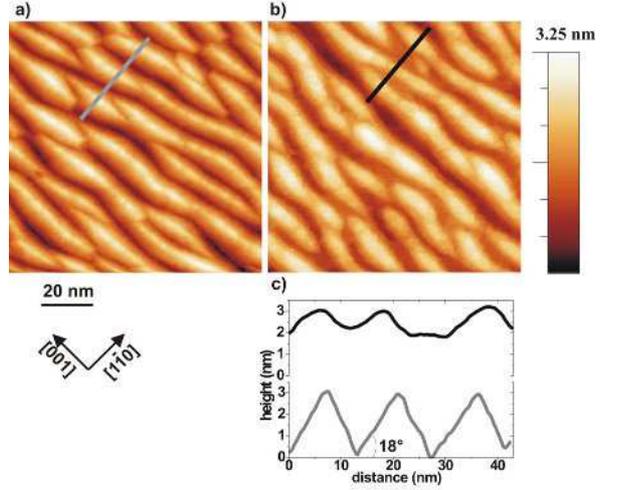}
    \caption{(a,b) $\unit[100]{nm}\times\unit[100]{nm}$ STM images  of (a)~the W(110) template (b)~$\unit[\approx1]{nm}$-thick wires
    of Fe($\Theta=\unit[2.5]{AL}$)/W. (c) Cross-sections along the lines shown in (a)-gray and in (b)-black}
    \label{fig:Graphique1}
\end{figure}
The ordered arrays of nanowires are obtained in several steps, described below.

\begin{figure*}
  \center
   \includegraphics[width=\linewidth]{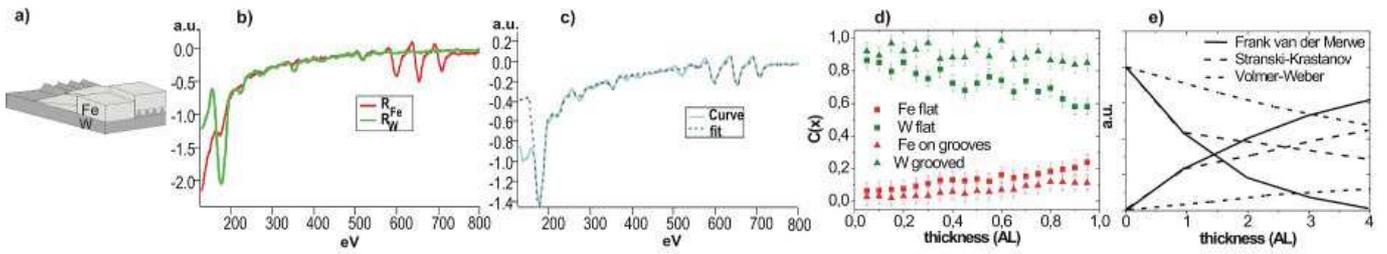}
    \caption{a) Schematic view of the sample prepared for the quantitative AES analysis, starting from two areas with a flat and a grooved W(110) surface. The two ends provide thick layers of W and Fe, connected by a continuous wedge of Fe b) Reference spectra of Fe ($R_{\mathrm{Fe}}$) and of W ($R_{\mathrm{W}}$) respectively c) Spectra at position $x$ along the wedge and the fit as a linear combination of the reference spectra d) Variation of the fitted intensities $c_{\mathrm{W}}$ and $c_{\mathrm{Fe}}$ of the W $NOO$ and Fe $LMM$ peaks over the wedge, for both the flat and the grooved surfaces e) Schematic representation of the expected curves for a Frank van der Merwe, Stranski-Krastanov and Volmer Weber growth.}
    \label{fig:figure}
\end{figure*}

The first step is the preparation of a smooth metallic buffer layer on sapphire
(11$\overline{2}$0) wafers. A seed layer of Mo (nominal thickness $\Theta=\unit[1]{nm}$) followed
by W ($\Theta\approx\unit[10]{nm}$) are deposited at room temperature (RT) followed by annealing
at \unit[800]{$^\circ$C}. This yields a smooth $(110)$ surface, with atomically-flat terraces of
mean width $\approx\unit[200]{nm}$\cite{Fru,Fru98}.

The second step consists of the preparation of the non-magnetic self-organized template. This step was inspired by reports of a
kinetic uniaxial roughening of films of the bcc elements Fe(110)\cite{Alb93} and W(110)\cite{Koh00} during
homoepitaxy at moderate temperatures. This roughening was explained by anisotropic diffusion along atomic steps
and the occurrence of an Ehrlich-Schwoebel barrier. Based on these clues we optimized the deposition parameters to produce self-organized arrays
of trenches aligned along the $[001]$ direction of the W(110) surface~(Fig.\ref{fig:Graphique1}a). Here we focus on arrays deposited at \unit[550]{$^\circ$C} for the first
atomic layers, and progressively reduced to \unit[150]{$^\circ$C} during growth. A stable period is then reached
within a few nanometers of deposition. The period and depth of the trenches were determined by STM to be
$\unit[\approx15]{nm}$ and $\unit[\approx2.5]{nm}$ for these deposition parameters, respectively. The trenches have well-defined facets of $\{210\}$ type, revealed by STM and RHEED
(tilted streaks) with a constant angle of $\theta\approx$\unit[18]{$^\circ$} with respect to the mean
surface\cite{borca, borca1}. The occurrence of facets with a well-defined angle minimizes the effect of the
fluctuations of the width of the trenches on the shape of the trenches, which thus displays essentially no
distribution. This is desirable as the distribution of magnetic anisotropy energy may increase tremendously for
nanosized systems in relation with the distribution of structural environments \cite{Roh06}.

\begin{figure*}
  \center
   \includegraphics[width=\linewidth]{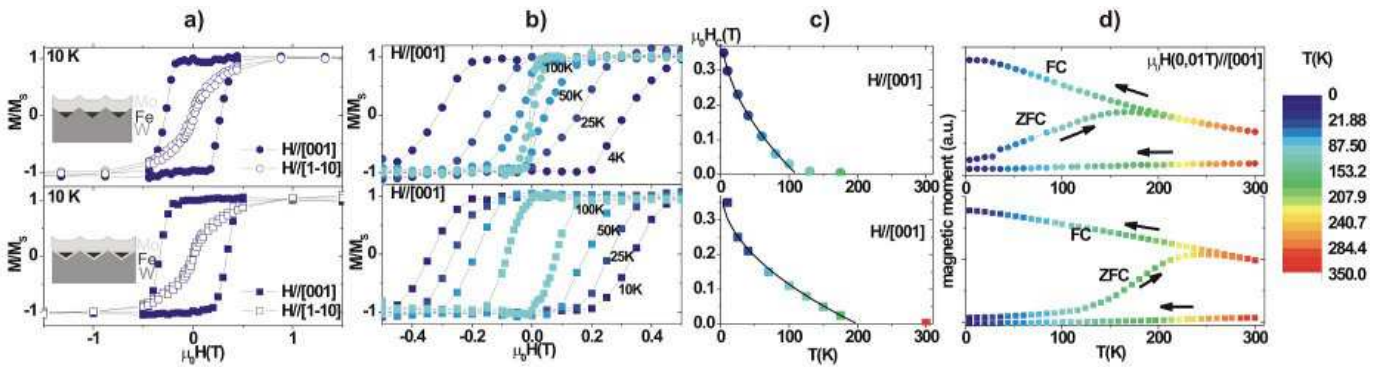}
    \caption{Magnetic measurements of Fe/W(110) wires with a capping layer of Mo; upper-part
    without under-layer, lower-part with Mo under-layer. a) Magnetization curves at \unit[10]{K}
    in-plane along [001] and [1$\overline{1}$0]. b) Magnetization curves in-plane along [001] at different
    temperature. c) Dependance of coercivity with temperature, along with a $\sqrt{T}$ fit. d)
    Zero-field-cooled (ZFC)/field-cooled (FC) curves, measured with a low magnetic field (\unit[0.01]{T}) applied
    along the in-plane [001] direction.} \label{fig:Graph2}
\end{figure*}

As an optional third step one atomic layer of Mo can be deposited at \unit[150]{$^\circ$C} on this
template, allowing a control of the chemical nature of its free surface without affecting the
morphology of the template nor the angle of the facets.

\begingroup

\def\cW{c_{\mathrm{W}}}%
\def\cFe{c_{\mathrm{Fe}}}%
\def\RW{R_{\mathrm{W}}}%
\def\RFe{R_{\mathrm{Fe}}}%

As the next step these self-organized templates are used for fabrication of magnetic wires. To this end we deposit Fe at
$150^\circ$C, a temperature under which Fe grows essentially layer-by-layer on \textsl{nominally-flat} W(110)\cite{Alb93}. STM reveals a leveling of the bottom of the trenches, suggesting their progressive filling and thus presumably yielding Fe wires with a
triangular cross-section~(Fig.\ref{fig:Graphique1}b-c). This conclusion is supported by the quantitative
analysis of AES spectra. To avoid errors in thickness resulting from fluctuations in deposition rates from one sample to another, all measurements were performed on one single wedged sample as depicted in Fig.\ref{fig:figure}a. Two ends of the sample consist of a \unit[3]{nm} thick Fe layer and of a \unit[9]{nm} thick W layer respectively, which provide reference spectra $R_{\mathrm{Fe}}$ and $R_{\mathrm{W}}$ for pure elements as well as a measure of the beam intensity (Fig.\ref{fig:figure}b). In-between these two areas Fe is deposited with a uniformly-varying thickness, both on a grooved and on a flat W(110) surface. Spectra at position $x$ along the wedge were fitted as a linear combination of the reference spectra (Fig.\ref{fig:figure}c) :

\begin{equation}
\label{eq-linearFitAES}
I(x)=\cW(x)\RW+\cFe(x)\RFe
\end{equation}

where $\cW(x)$ and $\cFe(x)$ are the intensity of W and Fe. Fig.\ref{fig:figure}d shows the variation of $\cW(x)$ and $\cFe(x)$, on the flat
and the grooved parts, with the Fe thickness $x$ varied in the range \unit[0-1]{AL} (Atomic layers). First notice that the mean free path of electrons at the $MNN$ Fe peaks is higher than that at the $NOO$ W peaks, explaining a faster decrease of $\cW$ than the increase signal of $\cFe$. Then, is it clear that the variation of the $\cW(x)$ and  $\cFe(x)$ is faster for the flat surface, where Fe is known to grow layer-by-layer (Frank van der Merwe mode), than for the grooved template. This behavior, reminiscent of the Volmer-Weber growth mode (Fig.\ref{fig:figure}e) with three-dimensional islands on a bare substrate\cite{Bauer}, may suggest that Fe does not or not fully wet the microfacets of the W template and flows from the beginning to the bottom of the grooves to form wires.

\endgroup

For the magnetic measurements reported here the growth of the wires has been stopped at $\Theta=\unit[2.5]{AL}$, \ie before the percolation threshold. The mean full width and mean full height are then expected to be
\unit[7]{nm} and \unit[1.2]{nm}, respectively. Notice that owing to the expected triangular shape of the wires, the height averaged over a wire equals half its full height.
%In the following we will simply call this average value \textsl{height}.

The final step is the deposition at room temperature of various capping layers (Au, Al, Mo, Pd), with the aim of modifying the magnetic anisotropy. The thickness of this layer is typically $\Theta=$\unit[4-5]{nm}, to act also as a protection against oxydation or other adsorbate-induced contamination, except for Pd where
only an interface layer of thickness $\Theta=$\unit[4]{AL} is deposited to set the magnetic anisotropy, capped by \thicknm{4} of Au as a protection against contamination.

\section{Magnetic measurements and analysis}

\begin{figure}
  \center
  \includegraphics[width=77mm]{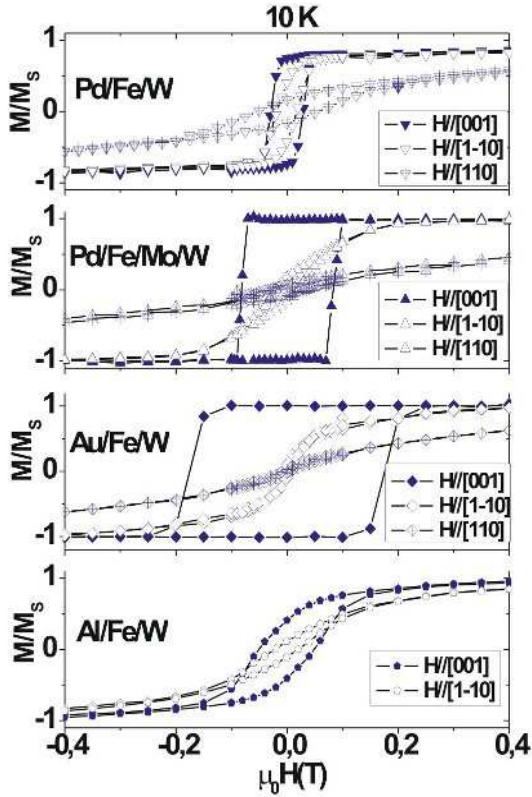}
   \caption{Magnetization curves at \unit[10]{K} in-plane along [001] and [1$\overline{1}$0], and perpendicular to the plane along [110] for Fe wires of type: a) Pd/Fe/W; b) Pd/Fe/Mo/W; c) Au/Fe/W; d) Al/Fe/W.}
    \label{fig:Graph3}
\end{figure}

The mean period, full width and full height of the wires investigated are \unit[15]{nm}, \unit[7]{nm} and \unit[1.2]{nm},
respectively. SQUID hysteresis loops were performed on
assemblies of such wires along three directions: along the direction normal to the plane, and along two in-plane
directions: parallel to $\mathrm{Fe}[001]$~(along the wires) and parallel to
$\mathrm{Fe}[1\overline{1}0]$~(across the wires).

We studied the influence on hysteresis loops of the chemical nature of both the top and of the
bottom interfaces. For the top interface the elements Pd, Au, Al and Mo have been
used~(Fig.\ref{fig:Graph3}). For the bottom interface only W and Mo have been used so as to keep
the growth process of the wires unchanged~(above room temperature Fe is liable to surface alloying
or intermixing when deposited on Al, Pd or Au). In all cases
the easy axis of magnetization essentially remains along $[001]$, \ie along the length of the wires. The
nature of both interfaces affects the coercivity along this direction, and the anisotropy fields
along the two harder directions, in-plane $[1-10]$ and out-of-plane $[110]$. The experimental
density of anisotropy energy along a hard direction was calculated from the hysteresis loop as
$E_{\mathrm{exp}}=\mu_{0}\int^{\Ms}_{0}H \diff{M}$. To minimize systematic errors related to any
small opening of the loop (caused \eg, by the misalignment of the sample in the Squid) we
processed the anhysteretic curve (sum of the two branches $M_\mathrm{up}$ and
$M_\mathrm{down}$) instead of one single branch. The experimental values thus determined of both
in-plane $K_\mathrm{tot,p}$ and out-of-plane $K_\mathrm{tot,\theta}$ densities of anisotropy energy are summarized in Tab.\ref{tab-Anisotropy}.
In the case of Pd/Fe/W hysteresis is observed in both in-plane directions, which reveals that the
anisotropy is weak and we are on the verge of a spin reorientation transition in-the-plane. In
this case a significant error probably arises from the crude procedure of using the calculated
anhysteretic curve for fitting the anisotropy.

\begin{table*}
    \centering
    \caption{Experimental in-plane (labeled~p) magnetic anisotropy energy density
    for various capping or underlayers,
    derived at \tempK{10} from the hysteresis loops of Fig.\ref{fig:Graph2} and Fig.\ref{fig:Graph3}. The numbers shown in brackets are
    theoretical ones, computed with values known for flat $(110)$ thin films \cite{Fru99,Fri94,Gradmann1993}.
    Concerning the differences only
    those numbers are displayed, which allow a comparison with a change of one interface only with Mo/Fe/W or Mo/Fe/Mo/W. Out-of-plane anisotropy has not been studied extensively, so that only the available figures are indicated.}
    \label{tab-Anisotropy}
        \begin{tabularx}{\linewidth}{l l X X l} %{*{5}{X}}%{r@{\hspace{3mm}}r@{\hspace{3mm}}r@{\hspace{3mm}}r@{\hspace{3mm}}r}
            \hline\hline
            Interfaces & $K_\mathrm{tot,p}$  ($\unit[10^5]{J/m^3}$)
                       & $\Delta K_\mathrm{s,p}$ ($\unit[]{mJ/m^2}$) with Mo/Fe/W
                       & $\Delta K_\mathrm{s,p}$ ($\unit[]{mJ/m^2}$) with Mo/Fe/Mo/W
                       & $K_\mathrm{tot,\theta}$ ($\unit[10^5]{J/m^3}$) \\
            \hline
            Mo/Fe/W    &   $3.6\pm0.6$ & --- & $-0.03\pm0.04(-0.56)$  \\

            Mo/Fe/Mo/W &   $4.2\pm0.4$ & $0.03\pm0.04(0.56)$ & ---  \\

            Pd/Fe/Mo/W &   $1.2\pm0.2$ & $-0.12\pm0.03(0.18)$ & $-0.15\pm0.02(-0.38)$ & $16.3\pm0.06$  \\

            Pd/Fe/W    &   $0.35\pm0.15$ & $-0.16\pm0.03(-0.38)$ & --- & $7.1\pm0.1$ \\

            Au/Fe/W    &   $1.7\pm0.3$ & $-0.10\pm0.03(0.21)$ & --- & $7.4\pm0.1$ \\
            Al/Fe/W    &   $0.9\pm0.05$ &  $-0.14\pm0.03(-0.24)$ & --- \\
            \hline\hline
        \end{tabularx}
\end{table*}

Let us analyze these values, with a final discussion on the magnitude of surface anisotropy with respect to
values known for flat thin films. The total anisotropy originates as the sum of several contributions. Some of
them can be estimated, while others are entangled and their distinction is mainly a view of the mind. We use the
following indices to label anisotropy constants and demagnetizing factors: no index for the total energy, d for
dipolar energy (both expressed in $\unit[]{J/m^3}$), s~for surface (expressed in $\unit[]{J/m^2}$), p~for
in-plane~(all with the convention that positive figures mean easy axis along $[001]$), $\theta$ for out-of-plane
(with the convention that negative means easy axis perpendicular to the plane). Notice that unlike some of the existing
works \cite{Elmers1990,Gradmann1993} we use the same sign convention for all anisotropies (total, volume and
surface).

\def\NpWire{N_\mathrm{W,p}}% Coefficient de champ d\'{e}magn\'{e}tisant de fil seul, transverse
\def\NthetaWire{N_{\mathrm{W},\theta}}% Coefficient de champ d\'{e}magn\'{e}tisant de fil seul, perpendiculaire
\def\NpArray{N_\mathrm{A,p}}% Coefficient de champ d\'{e}magn\'{e}tisant de r\'{e}seau complet, transverse
\def\NthetaArray{N_{\mathrm{A},\theta}}% Coefficient de champ d\'{e}magn\'{e}tisant de r\'{e}seau complet, perpendiculaire

The first type of anisotropy is the bulk fourth-order anisotropy of Fe, with
$K_1=\unit[\scientific{4.58}{4}]{J/m^3}$. When projected along a plane of the $(<110>,<001>)$ family this term
yields a second order contribution $K_1$ and a fourth order contribution $-(3/4)K_1$ \cite{Gradmann1993}, whose
sum is negligible before most figures found experimentally here.

The second type is dipolar energy, arising from
both the self energy of a wire, and from the interactions within the array. With the idealized picture of
equally-spaced infinite cylinders with a triangular isosceles base, let us call $\NpWire$ and $\NthetaWire$ the transverse and perpendicular demagnetization coefficients of an isolated wire, and $\NpArray$ and $\NthetaArray$ the
coefficients for the entire array (also including the self-dipolar energy of each wire). $\NpWire+\NthetaWire=1$ and $\NpArray+\NthetaArray=1$ because the wires are assumed to be infinitely long in the $[001]$ direction. Based on analytical calculations one finds $\NpWire\simeq0.147$~(see Annex~1) with the present geometry. The effect of the array is only minor and one finds, again analytically(Annex~2): $\NpArray\simeq0.142$. Based on bulk Fe magnetization $\Ms=\unit[\scientific{1.73}{6}]{A/m}$ demagnetizing energy densities are:
$K_\mathrm{d,W,p}=\unit[\scientific{2.76}{5}]{J/m^3}$ and $K_\mathrm{d,W,\theta}=\unit[\scientific{1.60}{6}]{J/m^3}$, and $K_\mathrm{d,A,p}=\unit[\scientific{2.67}{5}]{J/m^3}$ and $K_\mathrm{d,A,\theta}=\unit[\scientific{1.61}{6}]{J/m^3}$. Thus dipolar energy arises primarily from self-dipolar energy of individual wires, and is in magnitude and sign a major contribution to the value measured experimentally.

Another contribution to magnetic anisotropy is of magneto-elastic origin, both linear and non-linear \cite{Sander99}.
However unlike thin films investigated so far, stress is both in-plane and out-of plane owing to the occurrence
of many steps, implying a significant shear because W and Fe have a $\unit[10]{\%}$ lattice mismatch. Therefore
no figure or even sign can be reliably predicted, as no reference figure are available for the non-linear corrections to magneto-elasticity.

All three types of anisotropies mentioned above are of volume
type and to first approximation should be identical for all systems reported here, of identical thickness. With this respect, notice
that the $\unit[1]{AL}$-thick Mo underlayer is pseudomorphic with W so that no difference in magneto-elastic
contribution is expected with or without the underlayer. Then there remains surface anisotropies, which should
thus be responsible for any change found between the different samples. Whereas absolute values of interfacial
anisotropy can not be derived for wires, differences of interfacial energy between wires where only
the top interface is changed, can be derived from these experiments\brackettabref{tab-Anisotropy}. These differences are discussed in the following.

\begin{figure}
\center
  \includegraphics[width=77mm]{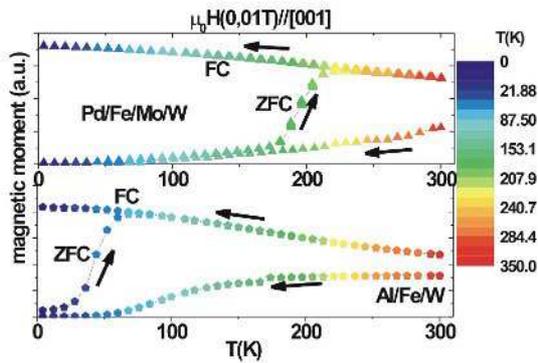}
   \caption{ZFC/FC curves measured with a low magnetic field (\unit[0.01]{T}) applied along the [001] direction for: a) Pd/Fe/Mo/W; b) Al/Fe/W.}
   \label{fig:Graph4}
    %\label{fig:Measy}
    %\label{fig:MT}
\end{figure}

Let us first notice the consistency within the present various experimental data. Experimental figures are available
for the change from Mo to Pd capping interface for both W and Mo underlayer. The resulting changes of surface
anisotropy are in very good agreement one with another, $\unit[-0.16]{mJ/m^2}$ and $\unit[-0.15]{mJ/m^2}$, respectively. This
noticed, let us discuss the results with respect to the literature. The change of magnetic anisotropy upon
changing the interface, computed from published data concerning thin films, are shown in brackets in
\tabref{tab-Anisotropy}. The experimental~(wires) and computed~(films) figures concerning the change of the bottom
interface~(first three lines of \tabref{tab-Anisotropy}) are in clear disagreement. This may point at the role
of the atomic steps, which are known to induce an extra anisotropy energy per atom with respect to a flat
surface\cite{Albrecht92, Rusponi03}. The step-edge anisotropy for Fe/W(110) with steps oriented along the [001]
direction was determined to be $K_{\mathrm{step,p}}\approx\unit[-1\times10^{-13}]{J/m}$\cite{Albrecht92}. With
one step every other $\unit[6.7]{\AA}$ for $\{210\}$ facets, this yields $\unit[0.2]{mJ/m^2}$ as an order of
magnitude for the extra contribution to the interfacial anisotropy energy. This figure is of the same order of magnitude as the discrepancy
between our experimental data for the stepped underlayer interface and the data computed from flat thin films.
However no experimental data is available for vicinal Mo surfaces so that the discussion cannot be pushed forward. The
figures for the change of anisotropy energy related to cappings layers are also in disagreement (see especially the case of Au), although no change of step anisotropy is expected in this case. This may hint to a dependance of surface anisotropy with strain, here both in-plane and out-of-plane, once claimed\cite{Bochi96} and since revisited in terms of non-linear magnetoelastic coupling coefficients in the simple situation of uniform strain\cite{Loser2000}.

Concerning the out-of-plane anisotropy, experimental figures are available for Pd and Au capping layers. A very small difference in the experimental values of the magnetic anisotropy is obtained for Pd/Fe/W and Au/Fe/W systems (\tabref{tab-Anisotropy}). This feature is in good agreement with that of the thin films, contrary to the case of in-plane anisotropy (discussed previously). With the same reassessment as for the in-plane anisotropy and considering the known data from thin films for $K_\mathrm{s,\theta}$(Au/Fe)$=-\unit[0.72]{mJ/m^2}$  and $K_\mathrm{s,\theta}$(Pd/Fe)$=-\unit[0.66]{mJ/m^2}$   \cite{Gradmann1993} it is figured out that magneto-elastic terms and deformations favor the out-of-plane configuration with an anisotropy energy of $\approx\unit[2.2\times10^{6}]{J/m^3}$. The observed behavior points at the dependence of the surface anisotropy with the strain. In this sense, the STM and RHEED measurements (not presented here), because no dislocations where observed, suggest that the wires are pseudomorphic up to thicknesses around \unit[1]{nm} higher than that of the thin films ($<\unit[2]{ALs}$). This has probably a major impact on magneto-elastic anisotropy.

 \begin{table*}
 \caption{The blocking temperature $\Tb$ determined from equation (3) and (4) for $\Hc(\Tb)=0$ and in brackets  from ZFC curves respectively, the nucleation volume $\Vn$ and the corresponding nucleation length for various capping or underlayers, and the domain wall width}
  \label{tab-Tb}
        \begin{tabularx}{\linewidth}{*{5}{X}}%{r@{\hspace{3mm}}r@{\hspace{3mm}}r@{\hspace{3mm}}r@{\hspace{3mm}}r}
            \hline\hline
            Interfaces        &   $\Tb$(K)&   $\Vn$(nm$^3$) &  $l$(nm)& wall width (nm)\\
            \hline
            Mo/Fe/W       &   110 (100) & 75 (68) & 18 (16) & 23 \\

            Mo/Fe/Mo    &   200 (190)& 165 (157) & 39 (37) & 22  \\

            Pd/Fe/Mo    &   (205)& (590)& (140) & 41 \\

            Al/Fe/W       &   (50) & (190) & (45) & 47 \\
            \hline\hline
        \end{tabularx}
\end{table*}

We finally discuss some aspects of coercivity, which unlike anisotropy is an extrinsic phenomena and as such may
yield information on \eg, the microstructure of a system. We focus on the case of Mo-capped Fe/W wires. At low temperature the hysteresis loops measured with the external field $H//[001]$ have a rather square shape, and at remanence full saturation is
still observed\bracketsubfigref{fig:Graph2}{b}. The coercivity continuously decreases with temperature while
$\Ms$ remains essentially unchanged suggesting a superparamagnetic behavior when coercivity vanishes. The
temperature-dependent coercivity $\Hc(T)$ is plotted in \subfigref{fig:Graph2}{c}. The decrease of $\Hc(T)$
reasonably follows a $\sqrt{T}$ law, as expected for a uniaxial anisotropy $K$ within the framework of the
coherent rotation model, assuming in a first approximation that $K$ does not depend on temperature. Under these two hypotheses the energy barrier preventing the reversal of magnetization
varies like

\begin{equation}
  \Delta E\propto [1-H/\Hc(T=\tempK{0})]^2,
\end{equation}

and assuming a thermally activated magnetization
switching with a switching rate following an Arrhenius law with a time constant

\begin{equation}
  \tau=\tau_{\mathrm{0}}\exp({\Delta E/ \mathrm{k}_{\mathrm{B}}}T),
\end{equation}

one derives:

\begin{equation}
  H_{\mathrm{c}}=\frac{2K}{\mu_{0}M_{S}}\left[1-\sqrt{\frac{\ln(\tau/ \tau_{\mathrm{0}})k_\mathrm{B}
T}{KV}}\right].
\end{equation}

The blocking temperature $\Tb$ determined as $\Hc(\Tb)=0$ is $\approx \unit[110]{K}$. This
agrees well with the half-way-up of the zero-field cooling remagnetization curve~(found at $\tempK{100}$), which
should represent the mean value of the blocking temperature. Using the relationship
$KV_{\mathrm{n}}\simeq25k_\mathrm{B}T_{\mathrm{B}}$ where $\Vn$ is the nucleation volume, inferred from an
attempt frequency $\tau_{\mathrm{0}}\approx\unit[10^{-10}]{s}$ and a measurement duration $\tau\approx\unit[10]{s}$,
we conclude that $V_{\mathrm{n}}\approx\unit[75]{nm^3}$, with the experimental anisotropy energy and the
$T_{\mathrm{B}}$ determined from the $H_{\mathrm{c}}(T)$ variation. Based on the estimated cross-section area of
the wires ($\unit[4.2]{nm^2}$), $\Vn$ is converted into a length of $\sim$\unit[18]{nm}. This length is
comparable to the domain wall width $\pi\sqrt{A/K}$=\unit[23]{nm}. This suggests that the Mo/Fe/W wires behave
essentially like individual wires. A further argument enforcing this conclusion is the progressive
remagnetization upon heating during the FC measurements. Both a much larger $\Vn$ and a sharp transition in the
rising ZFC curve occur in the case of strongly coupled wires \cite{Bor06}.

In the view of this analysis of coercivity the dependence of the magnetic anisotropy on the
underlayer and capping layer can be used to tune the blocking temperature. This is illustrated on
\figref{fig:Graph4} were systems with high and low anisotropy yield a change of mean blocking
temperature by a factor of four, with an identical volume.

\tabref{tab-Tb} summarizes the values of $\Tb$ determined from  $\Hc(\Tb)=0$ variation and/or ZFC-FC curves, and the resulting nucleation volume $\Vn$ for different available systems. The values of the length obtained from the conversion of the $\Vn$, remain comparable to the domain wall width for almost all the cases. Thus the wires are essentially isolated and have an individual magnetic behavior. An exception occurs for Pd/Fe/Mo/W, when the corresponding nucleation length  $\approx\unit[140]{nm}$ is much higher than the domain wall width. The coupling between the wires probably arises through the Pd layer, due to its high induced polarization, as observed in magnetic multilayers \cite{cheng,vogel}.

%\comment{J'ai enlev\'{e} toute discussion sur l'anisotropie perpendiculaire. On pourrait la faire,
%mais alors il faudrait faire les calculs sur les courbes de la constante d'anisotropie, comme tu
%l'avais fait pour les constantes planaires. Je te laisse voir si c'est faisable avec une certaine
%confiance.}

\section{Conclusion}

Arrays of epitaxial Fe(110) nanowires were obtained by deposition on kinetically self-organized
trenched surfaces of W, with a period of \unit[15]{nm}. Owing to a self-limiting effect the
nano-facets of the trenches are of type \{210\} with very little angular spread, which is suitable for reaching
a low distribution of physical properties such as magnetic anisotropy. Several underlayer~(Mo, W)
and capping layer~(Mo, Ag, Pd, Al) elements were used and the differences of in-plane magnetic
surface energy was measured. A significant disagreement was found with figures computed from values
determined on flat thin films, based on literature data. This points at the change of magnetic anisotropy owing to steps, already established, and possibly the dependance of surface anisotropy with strain, which is large
in such wires, both of tensile and shear type. In the course of the discussion of magnetic anisotropy, demagnetizing factors for infinite cylinders with an isosceles basis were calculated analytically, as well as the impact of the ordering as an array  on these coefficients.

\section*{Acknowledgments}

Two of the authors (B.~B. and A.~R.) acknowledge financial support from French R\'{e}gion Rh\^{o}ne-Alpes
(mobility program) and FP6 EU-NSF program (STRP 016447 MagDot), respectively. We are grateful to
V. Santonacci for technical support, and W. Wulfhekel for stimulating us for calculating the demagnetizing coefficients for cylinders with a triangular basis.

\section*{Annex~1: demagnetizing coefficients of a cylinder with triangular cross-section}

\begin{figure}
  \center
  \includegraphics{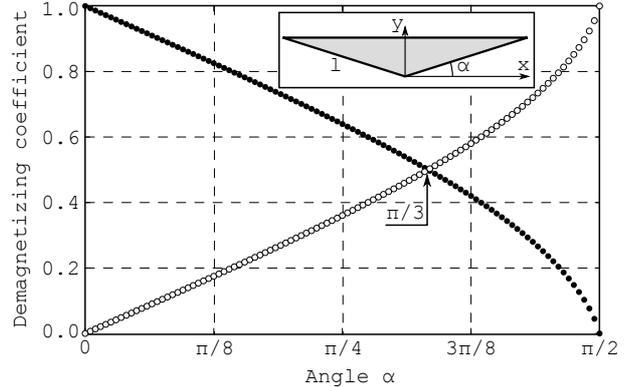}
    \caption{Variation of the demagnetizing coefficients $N_x$~($\circ$) and $N_y$~($\bullet$) with the base angle $\alpha$. The geometry of the triangle is depicted in the inset.}
   \label{fig-demag}
\end{figure}

In this section we calculate the demagnetizing coefficients $N_x$ and $N_y$ of an infinite cylinder with a triangular cross-section, such as $N_x+N_y=1$\footnote{With the notation of the experimental part: $N_x=\NpWire$ and $N_y=\NthetaWire$}. We do not consider a scalene triangle, however restrict the calculation to an isosceles triangle. Although demagnetizing fields are non-homogeneous and the self-consistency of the assumption of uniform magnetization can rigorously not be achieved under any arbitrary external field for volumes with surfaces not of second order, demagnetizing coefficients can still be defined via the dipolar energy of the hypothetically uniformly magnetized state \cite{Beleggia}. These coefficients are still related to the magnetization energy computed from hysteresis loops of infinitely-soft bodies\cite{Jubert1}. We describe the base isosceles triangle by the length $\ell$ of the two equal sides, and the angle $\alpha\in[0-\pi/2]$ between any of the two equal sides and the base of the triangle~(see inset of \figref{fig-demag}). In the following we use the shortcuts $c=\cos(\alpha)$ and $s=\sin(\alpha)$.

We evaluate the demagnetizing energy in the framework of charge and potential: $\Ed=(\muZero/2)\int \sigma\phi\:\diff S$, with $\vect H=-\Grad\phi$ and $\sigma=\vectMs\dotproduct\vect n$ ($\vect n$ is the outward normal to surfaces). The potential $\phi$ created at the origin by a horizontal linear segment of ends $i=1,2$ characterized by their ordinate $y$, abscissas $x_i$, radii $r_i$ and angles $\alpha_i$, and bearing a uniform charge $\sigma$, is

\begin{eqnarray}
  \label{eqn-potential-line}
  \nonumber\phi= & (\sigma/2\pi) \left[{x_2\log r_2 -x_1\log r_1 + \vphantom{y(\alpha_1-\alpha_2)}}\right. \\
        & \left.{y(\alpha_1-\alpha_2) + x_1-x_2}\right]
\end{eqnarray}

Assuming uniform magnetization along $x$ only the two equal sides of the triangle display magnetic charges, with a linear density $\sigma=s\Ms$. Making use of \eqnref{eqn-potential-line}, we find after some algebra:

\begin{equation}
  \label{eq-EdFinal}
  \Ed=\left({{\muZero \Ms^2 \over \pi}}\right)s^2\ell^2c(c\ln 2c + s\alpha).
\end{equation}

Upon normalizing by $(1/2)\muZero\Ms^2$ and by the area $\ell^2 cs$ of the triangle, we finally get the demagnetizing coefficients

\begin{eqnarray}
  \label{eq-NxNyFinal}
  N_x & = & 2s(c\ln 2c + s\alpha)/\pi \\
  N_y & = & 1-2s(c\ln 2c + s\alpha)/\pi.
\end{eqnarray}

\figref{fig-demag} shows the variation of $N_x$ and $N_y$ with $\alpha$. We may notice the values expected for limiting cases: $N_x(0)=0$ (thin-film with in-plane magnetization), $N_x(\pi/3)=1/2$ (equilateral triangle, \ie with vanishing two-fold anisotropy), $N_x(\pi/2)=1$ (thin-film with perpendicular magnetization).

One easily finds the expansions for flat triangles ($\alpha\rightarrow0$):

\begin{eqnarray}
  \label{eq-flatTriangle}
  N_x & = & {2\ln2 \over \pi} \alpha\\
   & = & {4\ln2 \over \pi}{t \over w} \\
   & \simeq & 0.88p
\end{eqnarray}

and for acute triangles ($\alpha \rightarrow \pi/2$):

\begin{eqnarray}
  \label{eq-acuteTriangle}
  N_y & = & -{2 \over \pi}
            \left({{\pi \over 2}-\alpha}\right)
            \ln\left({{\pi \over 2}-\alpha}\right) \\
      & = & -{1 \over \pi}\frac{w}{t}\ln \frac{w}{t} \\
      & \simeq & -0.32q\ln q
\end{eqnarray}

where $t=\ell s$ and $w=2\ell c$ are the full thickness (along $y$) and full width (along $x$) of the triangle, respectively, and $p=t/w\rightarrow0$ for $\alpha\rightarrow0$ and $q=w/t\rightarrow0$ for $\alpha\rightarrow\pi/2$ are the triangle full aspect ratios. These expansions might be compared to those of infinite cylinders with an elliptical cross-section with axes lengths $w$ and $t$: $N_x\sim p$ and $N_y\sim q$ for $\alpha\rightarrow0$ and $\alpha\rightarrow\pi/2$, respectively \cite{Osborn}. This shows that demagnetizing factors of infinite cylinders with an isosceles triangular basis are reasonably well described by an ellipsoidal basis with identical encompassing rectangle for a flat triangle, while it is not the case for an acute triangle.

\section*{Annex~2: demagnetizing coefficients of a 1D array of dipolar lines}

\begin{figure}
  \center
  \includegraphics{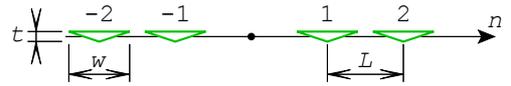}
    \caption{Geometry and naming for the calculation of the dipolar energy in an infinite chain of cylinders with an isosceles triangular cross-section.}
   \label{fig-infinitechain}
\end{figure}

In this section we estimate the corrections that shall be made to the demagnetizing coefficients of an isolated linear object such as the cylinders with triangular cross-section considered above, when an infinite one-dimensional array of them is now considered. We shall name $x$ the direction of the array, with period $L$\bracketfigref{fig-infinitechain}. The correction will be calculated under the hypothesis that the cylinders are sufficiently apart one from another so that they can be approximated with linear dipoles, \ie localized magnetic dipoles in a 2D space. Here $\mu=s \Ms$ where $s$ is the area of the cross-section of the cylinder, thus $s=wt/2$ here. Notice that in a 2D space the unit of magnetic dipoles is thus A.m.

It is readily shown that the stray field arising from a dipole $\vectmu$ in a 2D space is

\begin{equation}
  \vectHd(\vect r)=\frac{1}{2\pi r^2}\left[{2\frac{(\vectmu\dotproduct r)\vect r}{r^2}-\vectmu}\right]
\end{equation}

We proceed to calculation based on dipoles directed along the period of the array, so as to estimate the associated decrease of $N_x$: $\vectmu=\mu\vect i$. Noticing that $\sum_{n=1}^\infty(1/n^2)=\pi^2/6$, the stray field created at the origin by all other dipolar lines reads $(\mu\pi/6L^2)\vect i$. The correction to $N_x$ is thus:

\begin{equation}
  \Delta N_x=-\frac{\pi s}{6 L^2}
\end{equation}

In the case of cylinders with an isosceles cross-section, we have $\Delta N_x=-(\pi/12)(wt/L^2)$.

\end{document}